\documentclass[twocolumn,english,aps,prl,showpacs]{revtex4}
\usepackage[T1]{fontenc}
\usepackage{amsmath,graphicx,amssymb,epsfig,babel,dsfont,color}

\def\bbm[#1]{\mbox{\boldmath $#1$}}

\begin{document}

\title{Conduction-radiation coupling between two closely-separated solids}

\author{M. Reina}
\affiliation{Laboratoire Charles Fabry, UMR 8501, Institut d'Optique, CNRS, Universit\'{e} Paris-Saclay, 2 Avenue Augustin Fresnel, 91127 Palaiseau Cedex, France.}

\author{R. Messina}
\affiliation{Laboratoire Charles Fabry, UMR 8501, Institut d'Optique, CNRS, Universit\'{e} Paris-Saclay, 2 Avenue Augustin Fresnel, 91127 Palaiseau Cedex, France.}

\author{P. Ben-Abdallah}
\affiliation{Laboratoire Charles Fabry, UMR 8501, Institut d'Optique, CNRS, Universit\'{e} Paris-Saclay, 2 Avenue Augustin Fresnel, 91127 Palaiseau Cedex, France.}

\date{\today}


\begin{abstract}
In the theory of radiative heat exchanges between two closely-spaced bodies introduced by Polder and van Hove, no interplay between the heat carriers inside the materials and the photons crossing the separation gap is assumed. Here we release this constraint by developing a general theory to describe the conduction-radiation coupling between two solids of arbitrary size separated by a subwavelength separation gap. We show that, as a result of the temperature profile induced by the coupling with conduction, the radiative heat flux exchanged between two parallel slabs at nanometric distances can be several orders of magnitude smaller than the one predicted by the conventional theory. {These results could have important implications in the fields of nanoscale thermal management, near-field solid-state cooling and nanoscale energy conversion.}
\end{abstract}

\maketitle

Understanding the radiative heat transfer between two bodies at different temperatures is a very old problem in physics. At long separation distance, where energy exchange results exclusively from propagative photons, this transfer is well described by the radiometry theory introduced by Schuster~\cite{Schuster}, {which led} to the blackbody theory of Planck~\cite{Planck} at the begining of $20^{\text{th}}$ century. On the other hand, at subwavelength distances (i.e. in the near-field regime) the situation radically changes. Indeed, at this scale evanescent photons become the main contributors to the heat transfer by tunneling effect through the separation gap~\cite{Joulain}. {The basic foundations of heat transfer modeling at this scale have been laid} in the '70s with the work of Polder and van Hove (PvH)~\cite{Polder1973}, based on Rytov's theory of fluctuational electrodynamics~\cite{RytovBook1989}. In this semiclassical theory, the Poynting flux is calculated by suming up all the contributions generated by the random thermally-activated electric currents inside each body. This leads to the prediction of a dramatic amplification of radiative heat flux in the near field (with respect to the far field), which has been confirmed experimentally down to the nanometer range of distances considered in this work~\cite{Kittel,Reddy}.

\begin{figure}[hbt]
	 \centering
	 \includegraphics[width=0.4\textwidth]{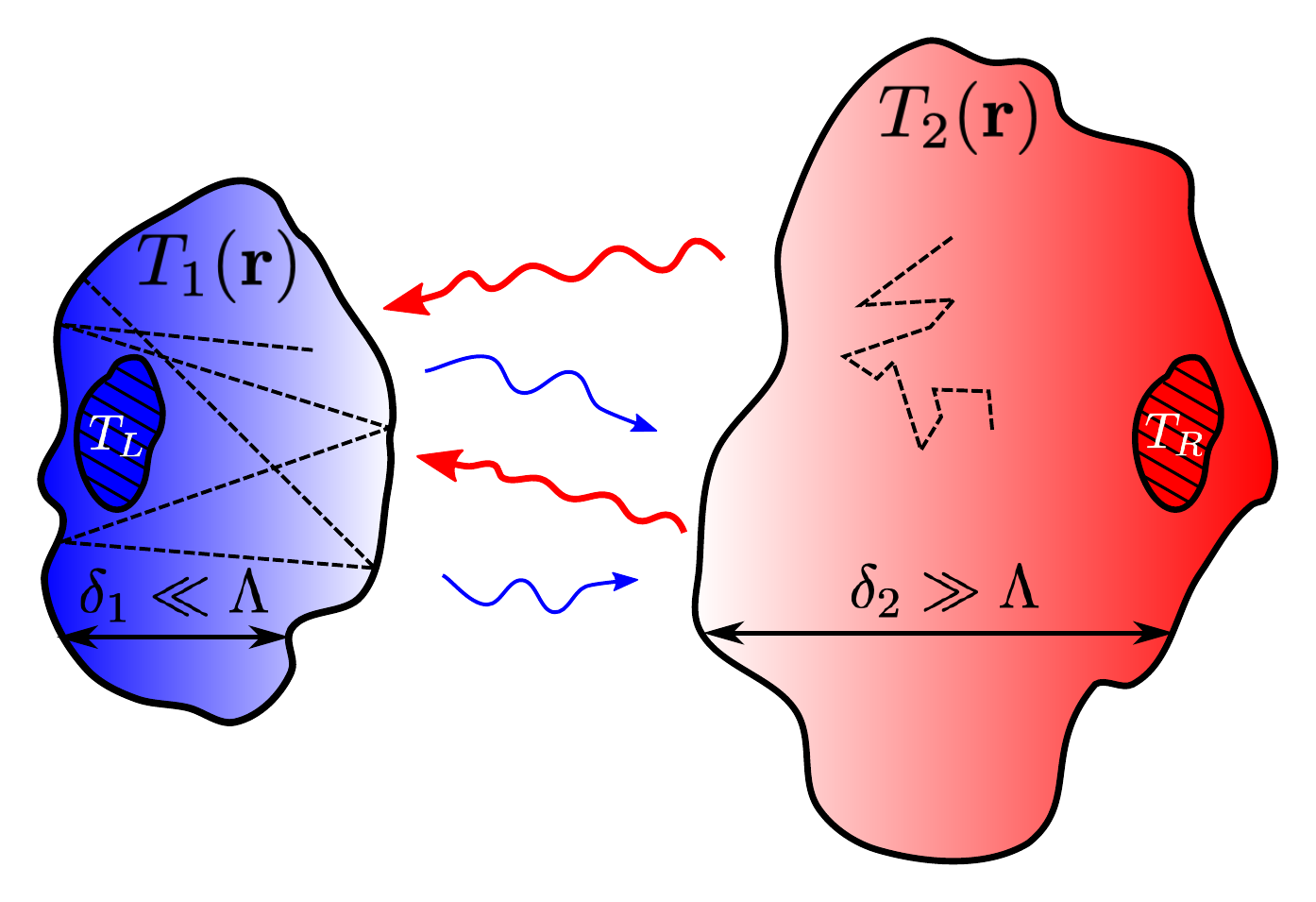}
	 \caption{Sketch of two bodies of finite size at different temperatures, partially coupled to two thermostats (hatched areas) at temperature $T_L$ and $T_R$, and exchanging heat radiatively through their separation gap. The black dashed lines show the heat-carrier (electron or phonon) trajectories between successive colliding events when the characteristic sizes $\delta_1$ and $\delta_2$ are respectively much smaller (i.e. ballistic regime) and much larger (i.e. diffusive regime) than the mean free path $\Lambda$. The temperature field $T_{1,2}(\mathbf{r})$ inside each body results from the local interplay between conduction and radiation.}
	 \label{Fig:system}
\end{figure}

{In the calculation of the net power exchanged between two solids held at uniform temperatures, the PvH theory neglects the coupling between the thermal photons which tunnel through the separation gap and the acoustic phonons inside each solid. Nevertheless, thermal photons are transported throughout each body and they dissipate their energy unevenly through them. Consequently, the temperature field within each body is generally not uniform and its spatio-temporal variation is driven by the conduction-radiation coupling between the two bodies}. A first attempt to describe this coupling has been proposed in 2016~\cite{Messina2016}. However, this phenomenological approach was limited to bodies of characteristic length much larger than the mean free path of heat carriers, so that no ballistic or partially ballistic transport could be taken into account.

In this Letter we introduce a general and self-consistent theoretical framework to describe the heat transfer between two solids of arbitrary size by taking into account the interplay between conduction and radiation. The essence of this approach is based on the combination of Boltzmann's equation to deal with the transport of heat carriers inside the solids (valid for any heat-transport regime) and fluctuational electrodynamics to calculate the radiative power locally dissipated in each body. {Our theory is limited to systems in the thermodynamic limit where the temperature is uniquely defined and where the local thermal equilibrium is reached. Moreover, the relative local temperature gradient is assumed to be small compared to the correlation length of the electromagnetic field inside the bodies~\cite{Eckhardt}. We also stress that we do not address here the problem of the transition between radiation and conduction, considered in some recent works~\cite{Chiloyan,Joulainb}.}

To start, let us consider two bodies as sketched in Fig.~\ref{Fig:system} assumed to be in partial contact with two thermostats and which are separated from each other by a subwavelength gap. We assume the thickness of this gap larger than the tunneling distance of electrons and acoustic phonons~\cite{Kittel,Reddy,Pendry,Prunnila}.
In these conditions, the internal energy density $u$ within these bodies obeys the conservation equation
\begin{equation}
	\frac{\partial u(\mathbf{r},t)}{\partial t}=P_{\text{rad}}(\mathbf{r},t)+P_{\text{cond}}(\mathbf{r},t),
	\label{Eq:energy_Eq}
\end{equation}
where $P_{\text{rad}}$ denotes the radiative power locally dissipated per unit volume within a given body and coming from the other one, while $P_{\text{cond}}$ is the conductive power per unit volume around the point $\mathbf{r}$, respectively. The latter can be calculated as the divergence of conductive flux
\begin{equation}
	\varphi_{\text{cond}}(t,\mathbf{r})=\sum_p\!\int_{4\pi} \!\!\!d\Omega\!\int\!d\omega\,\hbar \omega\, \mathbf{v}_{g,p}(\omega) f_p(t, \omega,\mathbf{r},\Omega)\frac{D_p(\omega)}{4\pi}, 
	\label{Eq:flux_cond}
\end{equation}
using the distribution function $f$ associated to the heat carriers within the solid, the density of states $D_p(\omega)$, the group velocity $\mathbf{v}_{g,p}(\omega)=\nabla_\mathbf{k}\omega_p$ of carriers at the frequency $\omega$ and solid angle $\Omega$. The distribution function $f_p$ for each polarization state $p$ can be calculated by solving Boltzmann's equation (for a given frequency $\omega$, not shown for simplicity) under the relaxation time approximation
\begin{equation}
	\frac{\partial f_p}{\partial t}+\mathbf{v}_{g,p}\cdot\nabla f_p=-\frac{f_p-f_0}{\tau_p(\omega,T(\mathbf{r}))},
	\label{Eq:Boltzmann_Eq}
\end{equation}
where $f_0$ is the equilibrium distribution (Fermi-Dirac for electrons and Bose-Einstein for phonons) and $\tau_p$ is the heat-carrier relaxation time.

\begin{figure}[hbt]
 \centering
 \includegraphics[width=0.47\textwidth]{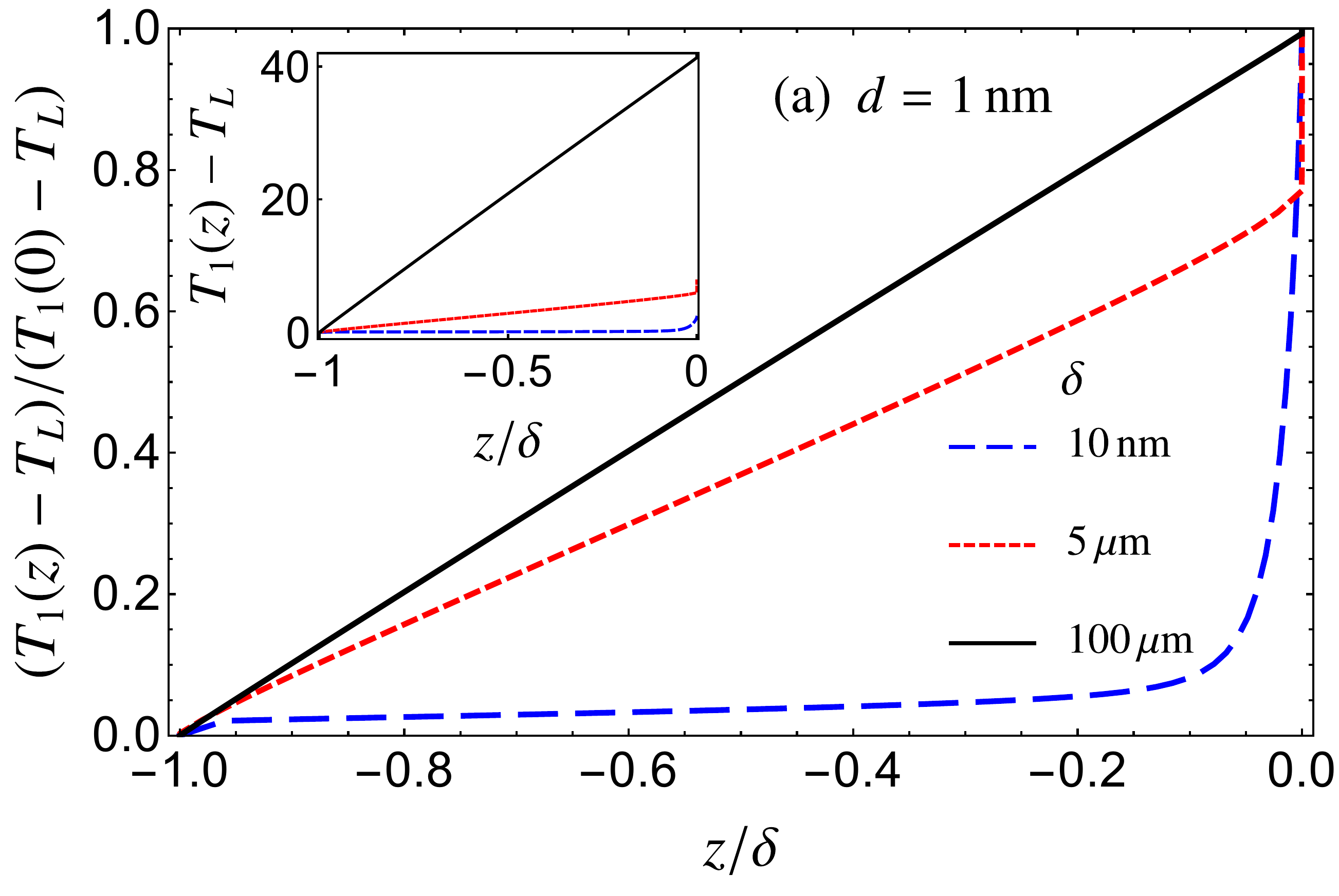}\\
 \includegraphics[width=0.47\textwidth]{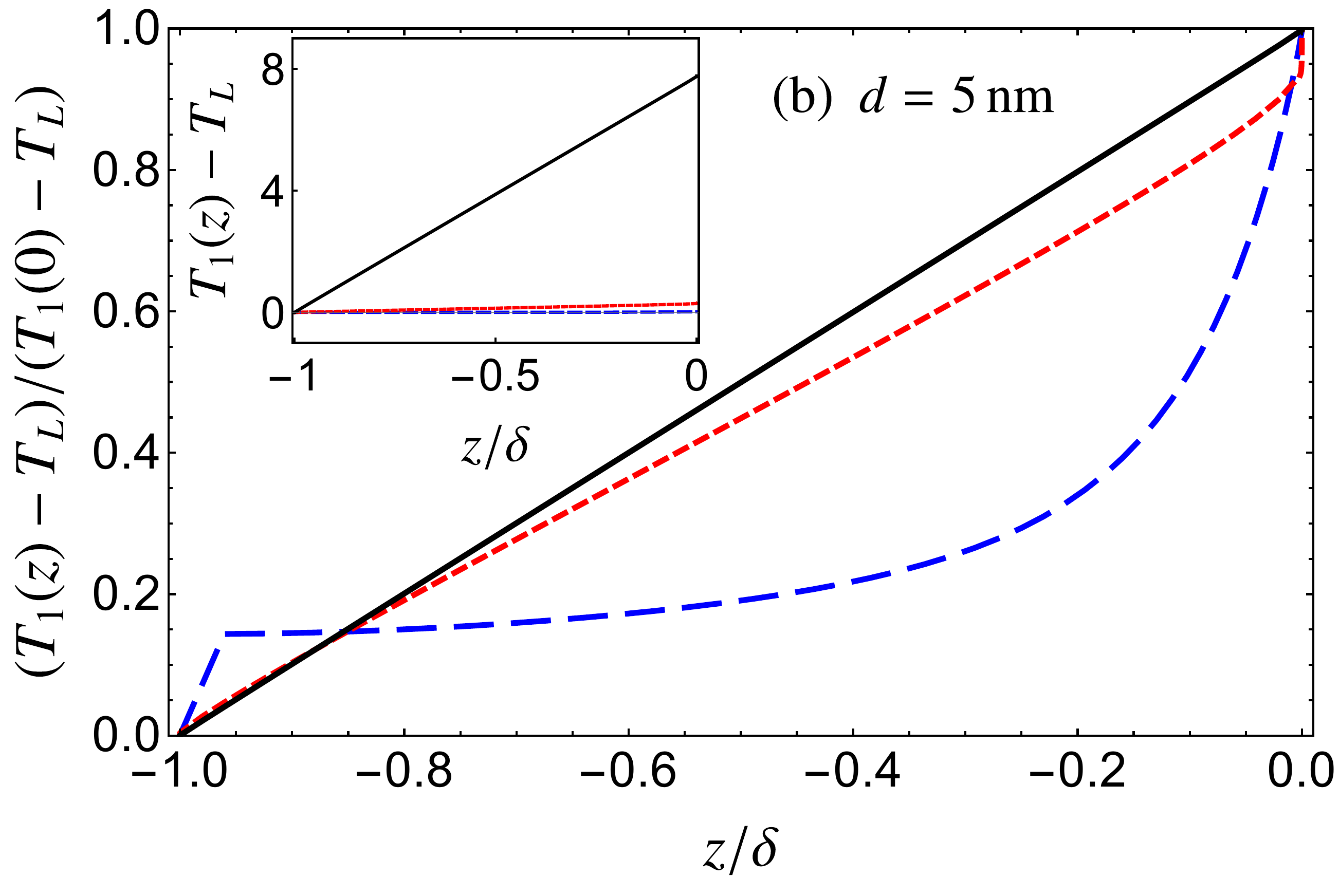}
 \caption{(a) Steady-state temperature (inset) and normalized temperature profile inside the left slab for different thicknesses and a separation distance $d=1\,$nm. (b) Same as (a) for $d=5\,$nm.}
\label{Fig_temp_profile}
\end{figure}

Concerning the radiative power, we start by neglecting the energy exchanged between parts of the same slab, assuming that this contribution is negligible with respect to conduction. The power $P^L_\text{rad}$ (resp. $P^R_\text{rad}$) dissipated in the left (resp. right) body and associated to the sources in the other body can be calculated from the net radiative flux $\bbm[\varphi]^R_\text{rad}$ (resp. $\bbm[\varphi]^L_\text{rad}$) using the statistical average $\langle\mathbf{S}(\mathbf{r},\omega)\rangle=2\,\text{Re}\langle \mathbf{E}(\mathbf{r},\omega)\times \mathbf{H^*}(\mathbf{r},\omega)\rangle$ of the Poynting vector spectrum at point $\mathbf{r}$ as
\begin{equation}
 	P^{L/R}_\text{rad}=-\int d\omega\,\nabla\cdot\bbm[\varphi]^{R/L}_\text{rad}(\mathbf{r},\omega).
	\label{Power_rad}
\end{equation}
According to the fluctuational-electrodynamics theory~\cite{RytovBook1989}, the contribution to the Poynting vector coming from the sources located in the left or right body {reads (using Einstein convention) for isotropic media when non-local effects are neglected}
\begin{equation}
\begin{split}
\langle S^{R,L}_{k}(\mathbf{r},\omega)\rangle&=i\frac{\omega^2}{c^2} \eta_{kjl}\int_{R,L}d\mathbf{r'} \epsilon''(\mathbf{r'},\omega)\Theta(T(\mathbf{r'}),\omega)\\
&\,\times[\mathds{G}^{EE}_{j,l}\mathds{G}^{EH*}_{k,l}-\mathds{G}^{EH*}_{j,l}\mathds{G}^{EE}_{k,l}],
\label{Poynting}
\end{split}
\end{equation}
where $\mathbf{r}$ is the point where the Poynting vector is calculated, while $\mathbf{r}'$ is evaluated in all points inside the source ($R$ or $L$). In Eq.~\eqref{Poynting}, $\eta_{kjl}$ are the components of Levi-Civita tensor ($k$, $j$ and $l$ referring to the three Cartesian coordinates), $\Theta(T,\omega)=\hbar\omega/[e^\frac{\hbar\omega}{k_BT}- 1]$ is the mean energy of a Planck oscillator at temperature $T$, $\epsilon''$ the imaginary part of the permittivity in the emitting body while $\mathds{G}^{EE}=\mathds{G}^{EE}(\mathbf{r},\mathbf{r'})$ and $\mathds{G}^{HE}=\mathds{G}^{HE}(\mathbf{r},\mathbf{r'})$ are the full electric-electric and electric-magnetic dyadic Green tensors at frequency $\omega$, taking into account all scattering events within the system between the emitter and the point where energy is dissipated~\cite{Joulain,Tomas}. {When calculating the monochromatic net radiative power (including both the power received by the other body and the one emitted by the body itself) appearing in $P_\text{rad}$ dissipated at position $\mathbf{r}$, we use Eq.~\eqref{Power_rad} [by taking the divergence of Eq.~\eqref{Poynting}] and finally replace $\Theta(T(\mathbf{r}'),\omega)$ by $\Theta(T(\mathbf{r}'),\omega)-\Theta(T(\mathbf{r}),\omega)$ in order to take into account the power emitted by the element located at $\mathbf{r}$ and ensure vanishing energy exchange at thermal equilibrium}.

To illustrate the importance of coupling mechanism between conduction and radiation in a two-body system in near-field interaction, we focus on a simple configuration made of two identical slabs of arbitrary thickness $\delta$ separated by a vacuum gap of thickness $d$, and in contact on their external sides with two thermostats at temperature $T_L$ and $T_R > T_L$. For the sake of clarity we consider slabs made of silicon carbide with a zincblende crystal structure (3C-SiC) and thicknesses larger than $10\,$nm, so that their dielectric permittivity{~\cite{Palik}} can be assumed to be size-independent. Using the dispersion relation of acoustic modes (giving the leading contribution to heat conduction), making the common isotropic assumption for wave vectors and considering the $[100]$ direction in the $\mathbf{k}$ space, we calculate~\cite{SM} the phonon relaxation time by taking into account the scattering by point impurities, the umklapp processes and the boundary scattering using Matthiessen's rule~\cite{Klemens}
\begin{equation}
	\tau^{-1}(\omega,T)=A\omega^4+B\omega^2T^3+C,
	\label{Eq:relaxation_time}
\end{equation}
\begin{widetext}
\hspace{0.cm}
\begin{figure}[hbt]
 \centering
 \includegraphics[width=0.47\textwidth]{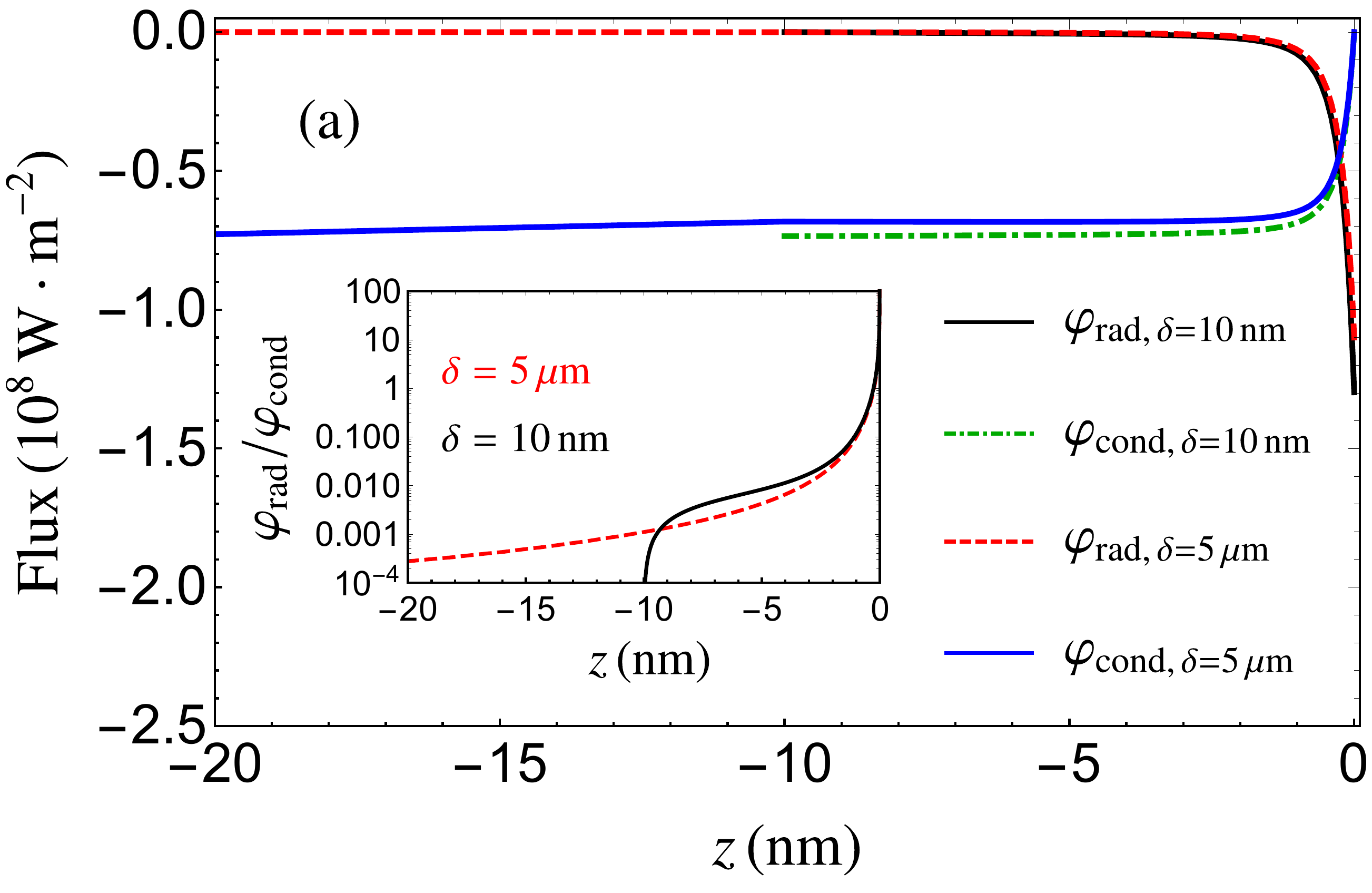}\hspace{.1cm}\includegraphics[width=0.47\textwidth]{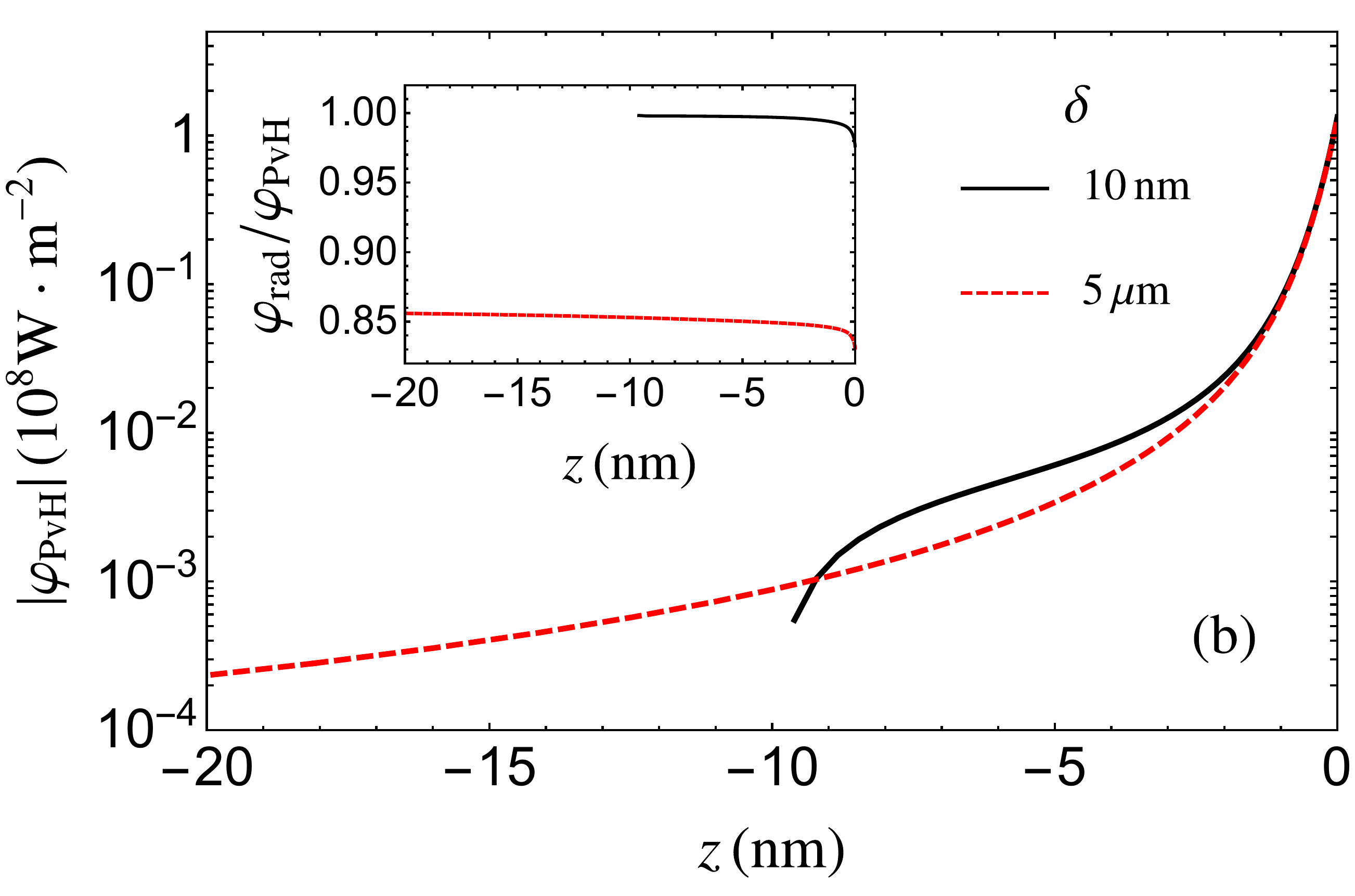}\\
 \includegraphics[width=0.47\textwidth]{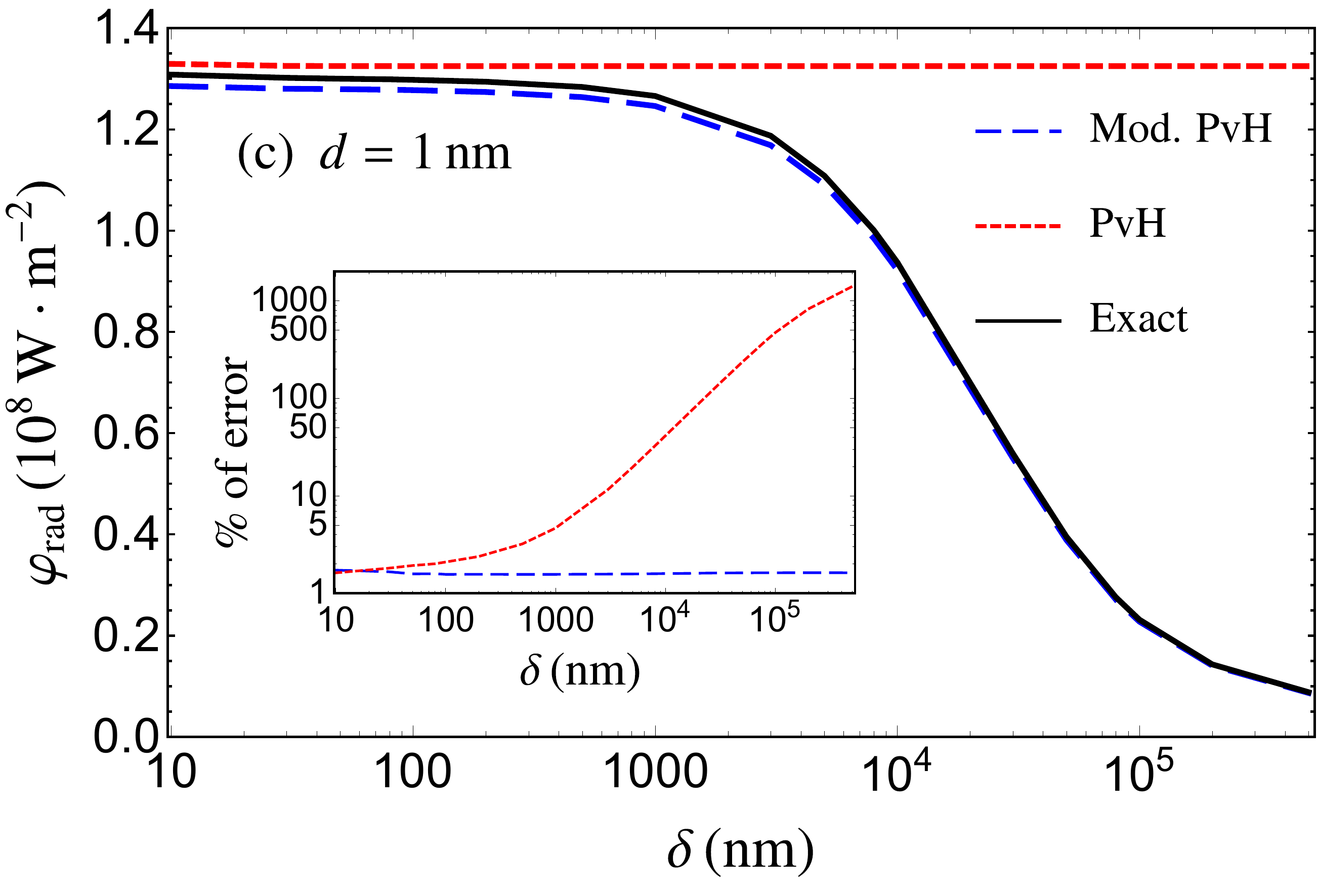}\hspace{.1cm}\includegraphics[width=0.47\textwidth]{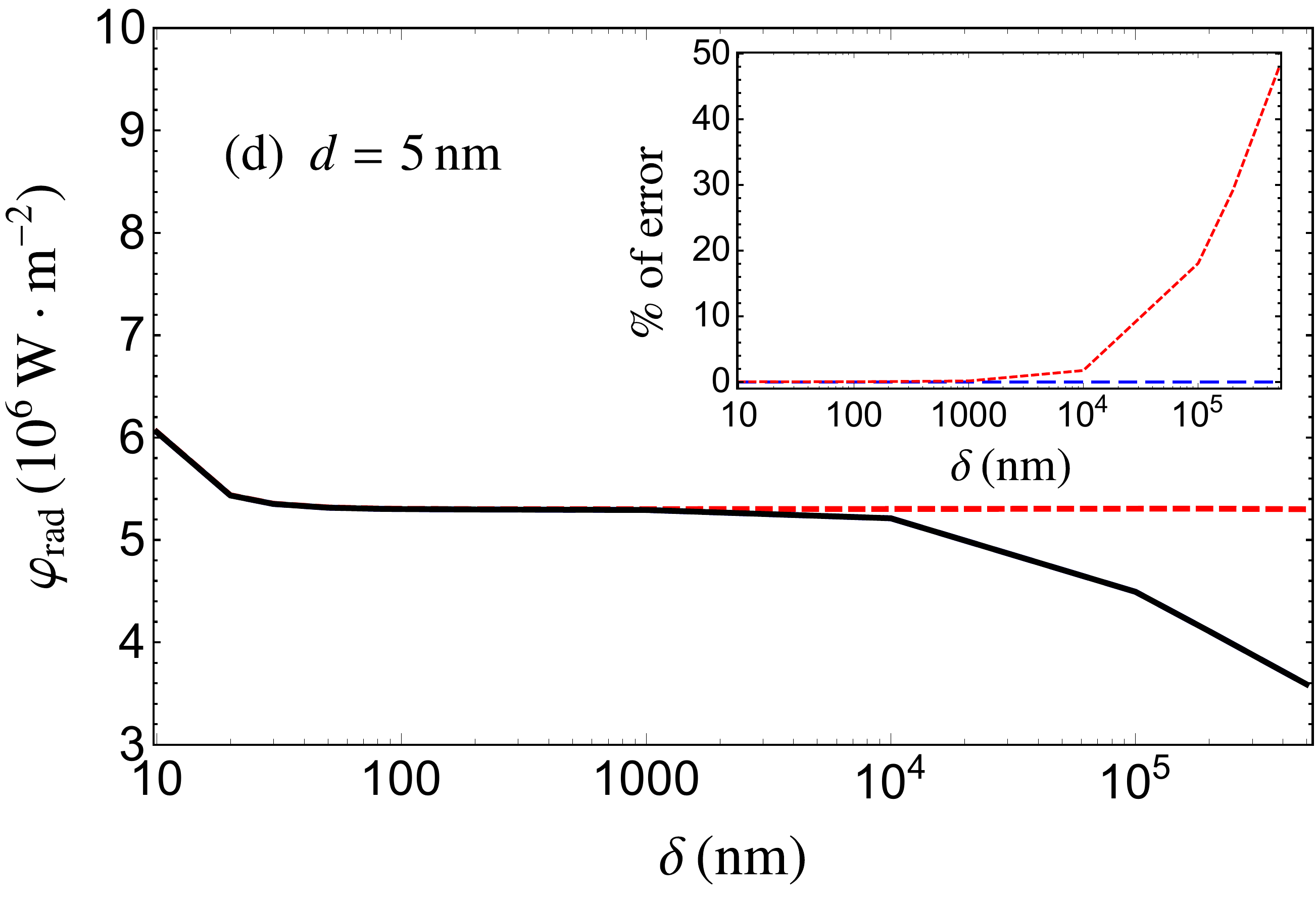}
 \caption{(a) Radiative flux $\varphi_{\text{rad}}$ within the left slab in a system of two 3C-SiC slabs of thickness $\delta=10\,$nm and $\delta=5\,\mu$m separared by a vacuum gap of thickness $d=1\,$nm and thermostated on their back sides at $T_L=300\,$K and $T_R=400\,$K. Inset: ratio between $\varphi_{\text{rad}}$ and the conductive flux $\varphi_{\text{cond}}$. (b) Absolute value of the PvH flux within the left slab for $\delta=10\,$nm and $\delta=5\,\mu$m for a separation distance $d=1\,$nm. Inset: ratio between the exact radiative flux $\varphi_{\text{rad}}$ and the PvH prediction. (c\.-\.d) Radiative heat flux exchanged between two 3C-SiC slabs with respect to their thickness for a separation distance of (a) $d=1\,$nm and (b) $d=5\,$nm. We show the exact result (black line), the PvH one (red dashed line, uniform temperatures $T_L=300\,$K and $T_R=400\,$K) and the modified PvH flux (blue long-dashed line, uniform temperatures equal to the temperatures at the boundaries with the vacuum gap in the steady states resulting from the coupling with conduction). Insets: absolute value of the error with respect to the PvH and modified PvH approaches.}
 \label{Fig_flux_vs_dist}
\end{figure}
\end{widetext}
where the coefficients $A= 2.1237 \times 10^{-45}\,\text{s}^3$, $B = 4.397\times 10^{-25}\,\text{s}\cdot\text{K}^{-3}$ and $C= 1.3949 \times 10^{8}\,\text{s}^{-1}$ have been obtained by fitting the simulated thermal conductivity~\cite{SM} with the available experimental data~\cite{Morelli} over the temperature range $[T_L,T_R]$. Using this expression we can derive the power dissipated by conduction from Eq.~\eqref{Eq:flux_cond}. As for the radiative power, it can be calculated from Eq.~\eqref{Power_rad} using the Green tensors in a multilayer geometry~\cite{Tomas}. By neglecting the contribution of propagative photons we obtain~\cite{SM}
\begin{equation}
	\begin{split}
			&P_{\text{rad}}(z)=\frac{2}{\pi^2}\sum_p\int_0^{+\infty}\!\!d\omega\int_{\frac{\omega}{c}}^{+\infty}\!\!dk \,k\, e^{-2\text{Im}(k_z)d}G_1(z,\omega)\\
			&\,\times\int_{0}^{\delta}dz'\Bigl(n\Bigl[\omega - \frac{eV_1}{\hbar}H(\omega-\omega_{g1}),T(z'+d)\Bigr]\\
			&\,\hspace{1cm}-n\Bigl[\omega - \frac{eV_2}{\hbar}H(\omega-\omega_{g2}),T(-z)\Bigr]\Bigr)G_2(z',\omega),
	\end{split}
\end{equation}
where $n(\omega,T)=[{e^{\frac{\hbar\omega}{k_B T}}-1}]^{-1}$ is the Bose-Einstein distribution function, {$H(x)$ the Heaviside step function}, while $G_1(z,\omega)$ and $G_2(z,\omega)$ are functions which depend on the optical properties of slabs~\cite{SM}. {This expression allows to compute the radiative power exchanged between two semiconductors with an applied voltage $V_i$ $(i=1,2)$, resulting in a modified photon statistics above their respective bandgap frequencies $\omega_{gi}$ (see~\cite{Cooling} for more details). }

The temperature profiles inside the slabs are obtained by solving through an iterative process Eq.~\eqref{Eq:energy_Eq} using the control angle discrete ordinates method~\cite{Ali} to solve Boltzmann's equation. For convenience, in the following we show temperature profiles in the left slabs, being the ones in the right slab qualitatively similar. The results in steady-state regime (i.e. for $\frac{\partial}{\partial t}\equiv 0$), are plotted in Fig.~\ref{Fig_temp_profile}(a) for different slab thicknesses and a separation distance $d=1\,$nm (normalized in the main part to compare the different profile shapes, in real values in the inset). When the thickness is small ($\delta=10\,$nm) compared to the mean free path of phonons~\cite{SM} the regime of transport becomes ballistic. It follows that the temperature profile becomes almost constant and close to the reservoir temperature $T_L$ (resp. $T_R$) in the left (resp. right) slab. Nevertheless, near the internal interfaces we note the presence of a sharp temperature variation. As shown in Fig.~\ref{Fig_flux_vs_dist}(a), this variation corresponds to the region where almost all the radiative energy carried by evanescent photons is deposited. This corresponds to the zone where the radiation-conduction coupling effectively takes place. As shown in the inset of Fig.~\ref{Fig_flux_vs_dist}(a) we see that for such thicknesses the radiative flux surpasses the conductive flux by two orders of magnitude close to the interface. Therefore, the phonons cannot cool down this region through their coupling with the external reservoir. As a result, the slab is significantly heated up locally (within some nm) close to the interface. On the other hand, beyond this region the conductive flux dominates the rapidly decaying radiative flux, so that the atomic lattice is thermalized at the reservoir temperture thanks to the ballistic phonons.
For thicker slabs ($\delta>5\,\mu$m), we see in the inset of Fig.~\ref{Fig_flux_vs_dist}(a) that the radiative flux still dominates over the conductive one within a few nm from the vacuum interface. However, in this case the regime of conduction tends to be diffusive and the atomic lattice does not thermalize anymore at the reservoir temperature. The temperature profile decays gradually (linearly for a purely diffusive regime) to the reservoir temperature thanks to the local colliding events of phonons. Figure \ref{Fig_temp_profile}(b) shows the results for a larger separation distance ($d=5\,$nm). While the overall qualitative behavior remains the same, the temperature drop is much smaller, due to the $1/d^2$ decay of the radiative flux. Also we note that the local radiative heating takes place at greater depth within the slab, the surface-parallel wave vectors of smaller value being preponderant for this separation distance.

We now want to address the impact of conduction/radiation coupling on the value of radiative flux. We first focus on the spatial distribution of radiative flux $\varphi_\text{rad}$ within the left slab. The results predicted by the PvH theory for two slabs set at uniform temperatures $T_L=300\,$K and $T_R=400\,$K are shown in Fig.~\ref{Fig_flux_vs_dist}(b) inside the first 20\,nm from the vacuum gap. For the three considered thicknesses the flux is rapidly decaying and its value is almost the same over the first 2\,nm~{\cite{Blandre}}. The inset shows the ratio between the exact value of the flux (taking into account the radiation/conduction coupling mechanism) and the PvH predictions. While for $\delta=10\,$nm the PvH description is reliable, for higher thicknesses it largely overestimates the exact flux, as a result of the conduction-induced temperature profile.

We finally focus on the net radiative flux exchanged between the two slabs and compare it to the flux predicted by the PvH theory when the two bodies are held at uniform temperature. More specifically, we compare the exact flux to the PvH one with $T_L=300\,$K and $T_R=400\,$K, and to the PvH result using as slab temperatures the values of the temperatures at the boundaries with the vacuum gap in the steady states derived from our approach. The latter is referred as \emph{modified PvH}. At $1\,$nm separation distance [Fig.\ref{Fig_flux_vs_dist}(c)], we see that for slab thicknesses larger than about $1\,\mu$m the discrepancy between the PvH prediction and our theory increases dramatically. The relative error is close to $5\%$ when $\delta=1\,\mu$m and scales as $\delta^2$ beyond this thickness. In slabs of such thicknesses the regime of heat transport becomes almost diffusive (see the phonon mean free path in~\cite{SM}) and the difference with the PvH theory comes from the linear variation of temperature profile which significantly reduces the temperature difference between the slabs. With thinner slabs the difference between the exact and the PvH theory becomes less pronounced, despite the temperature drop close to the internal interfaces highlighted previously. Nevertheless in these cases a relative error of about $2\%$ persists. Focusing on the modified PvH result, we note that it pretty well reproduces the exact results for any slab thickness. This demonstrates that the heat transfer between two solids in the extreme near field is mainly a surface-interaction mechanism. Nevertheless, while this is interesting from a fundamental point of view, we stress that the modified PvH calculation cannot be obtained without a full solution of the problem including the coupling mechanism. When the separation distance is increased to $d=5\,$nm we see [Fig.~\ref{Fig_flux_vs_dist}(d)] that for thin slabs (i.e. ballistic regime) the predictions of the PvH theory match perfectly well the exact calculation. In this case the radiative coupling between the two slabs is significantly smaller than at $d=1\,$nm, so that the induced temperature gradient is much smaller [see Fig.~\ref{Fig_temp_profile}(a)]. In this scenario, we only see a discrepancy with respect to the PvH results for large thicknesses, whereas the agreement with the modified PvH results is almost perfect. 
Moreover the comparison of results plotted in Figs.~\ref{Fig_flux_vs_dist}(c) and (d) shows that for thins films the radiative flux fits perfectly well the usual $1/d^2$ scaling law as predicted by the PvH theory. On the other hand, for thicker films (i.e. when the deviation with the PvH becomes more significant) this flux increases slower when the separation distance is reduced. This ``saturation or attenuation effect" induced by the radiation-conduction coupling is consistent with the previous observations~\cite{Messina2016}.

\begin{figure}[hbt]
 \centering
 \includegraphics[width=0.47\textwidth]{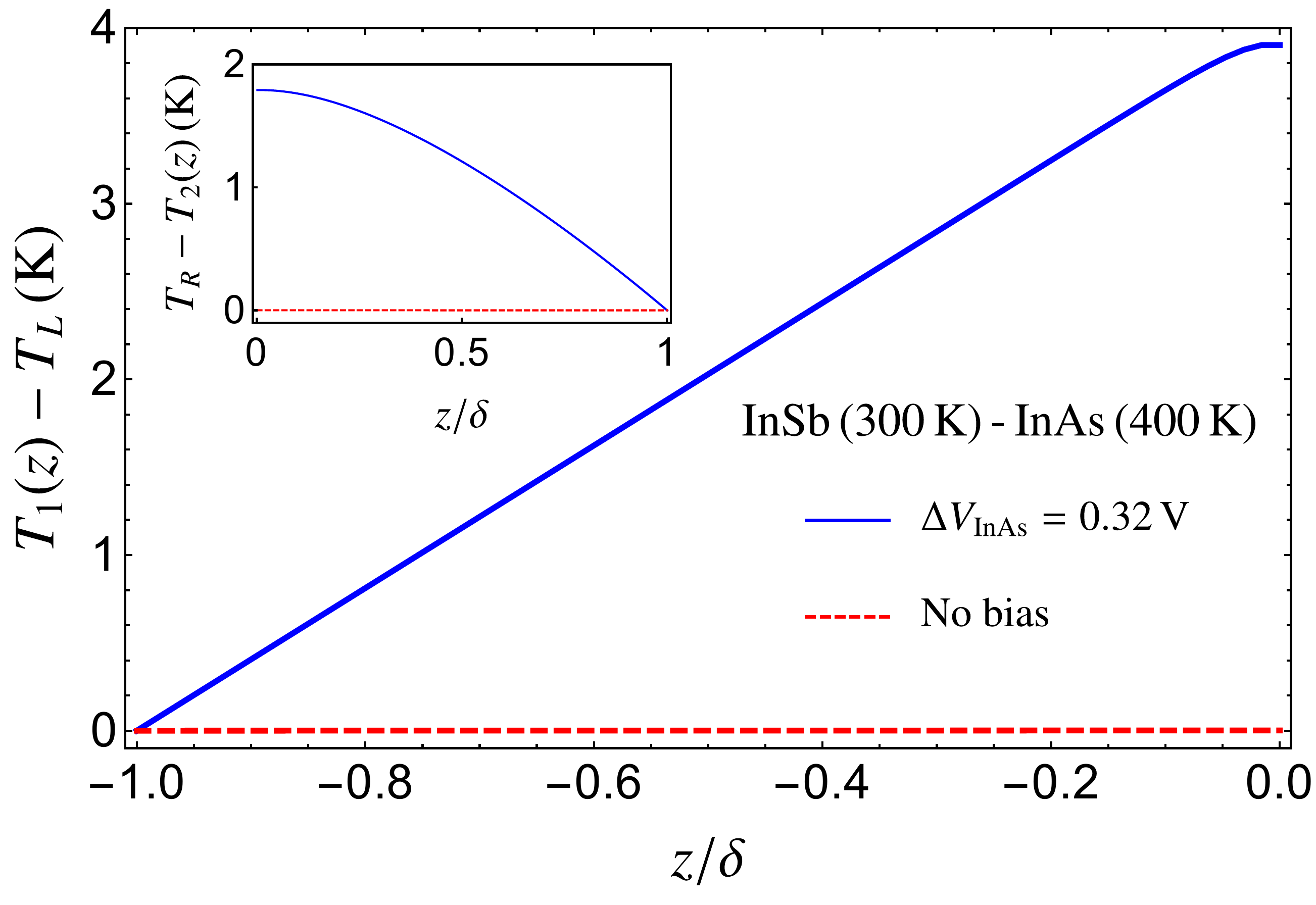}
 \caption{Temperature profile within a InSb slab (main part of the figure) $100\:\mu m$ thick at a distance $d=100\,$nm from a InAs slab of same thickness (inset), with or without an applied potential bias of $0.32\,$V. The external temperatures are $T_L=300\,$K and $T_R=400\,$K.}
 \label{Fig_bias}
\end{figure}

{Up to now we have applied our theory to systems made of solids with relatively high thermal conductivity. In weakly conducting solids the phonon-photon coupling and its thermal consequences can be radically different. This change is illustrated in Fig.~\ref{Fig_bias} for two thick slabs of semiconductors in interaction at $100\,$nm separation distance. In this case, when an external bias voltage is applied to the hotter body, the magnitude of radiative heat flux mediated by the evanescent photons is comparable to the heat flux carried by the acoustic phonons, and we clearly see that the temperature profiles dramatically differ from the ones (almost constant) obtained without bias. In this case and differently from the case of two polar materials (where the radiative heat exchange is mainly mediated by surface waves that is to say by localized resonant modes with large wavectors), the near-field heat exchanges beyond the semiconductor gaps come mainly from a continuum of frustrated modes which have by definition a small wavector. Hence in this case the radiative power is dissipated at the heart of solids even for relatively large separation distances.}
 
In conclusion, we have introduced a general theory to describe heat exchanges between two closely-spaced solids of arbitrary size. Our theory takes into account the conduction-radiation coupling between the two bodies, not included in PvH theory. By applying this theory to parallel planar slabs made of polar dielectrics {or semiconductors}, we have shown that this coupling produces an inhomogeneous temperature profile within each body, resulting in a radiative flux which can differ significantly from the one predicted by the PvH theory. {In weakly conducting semiconductors we have shown that the phonon-photon coupling  can dramatically modify the thermal state of solids up to separation gaps of hundred nanometers.  This theory can be relevant in the modelling of experiments exploring heat transfer in near-field and extreme near-field regime. It allows for a better temperature and heat-flux control at nanoscale and could find applications in the fields of thermal management, near-field solid-state cooling and nanoscale energy conversion.}

\begin{acknowledgements}

\end{acknowledgements}

\end{document}